\documentclass[11pt,letterpaper]{llncs}
\usepackage{amsmath}
\usepackage{amssymb}
\usepackage{graphicx}
\usepackage{marvosym}
\usepackage{url}
\usepackage{fancyhdr}

\usepackage{geometry}
\newcommand{\keywords}[1]{\par\addvspace\baselineskip
\noindent\keywordname\enspace\ignorespaces#1}

\fancypagestyle{firstpage}{\fancyhf{}
\setcounter{page}{69}
\fancyhead[C]{\scriptsize{International Journal of Computer Networks \& Communications (IJCNC) Vol.10, No.2, March 2018}}
\fancyfoot[L]{DOI: 10.5121/ijcnc.2018.10207}
\rfoot{\thepage}
}

\begin{document}


\title{\LARGE{Lightweight Cryptography for Distributed PKI Based MANETS}}


%
%
\author{\large{N Chaitanya Kumar \and Abdul Basit \and Priyadarshi Singh \and V. Ch. Venkaiah}}
\vspace{0.2cm}
\institute{\large{School of Computer and Information Sciences, University of Hyderabad,\\ Hyderabad-500046, India}}

%


%
%


\maketitle

\thispagestyle{firstpage}

\begin{abstract}
\textit{Because of lack of infrastructure and Central Authority(CA), secure communication is a challenging job in MANETs. 
A lightweight security solution is needed in MANET to balance its nodes resource tightness and mobility feature. The role of CA should be decentralized in MANET because the network is managed by the nodes themselves without any fixed infrastructure and centralized authority. In this paper, we created a distributed PUblic Key Infrastructure (PKI) using Shamir secret sharing mechanism which allows the nodes of the MANET to have a share of its private key. The traditional PKI protocols require centralized authority and heavy computing power to manage public and private keys, thus making them not suitable for MANETs. To establish a secure communication for the MANET nodes, we proposed a lightweight crypto protocol which requires limited resources, making it suitable for MANETs.}
\end{abstract}


\section{Introduction}

MANET known as Mobile ad hoc Network  is a self-configuring, a dynamic and infrastructure-less network of wireless connected 
mobile nodes \cite{anjum} \cite{wiki}. These mobile nodes move freely. Every node in the network has its own communication 
range and other nodes within the range can interact and exchange messages. If any nodes tend to go from the prescribed range of the MANET, it eventually leads to the failure of nodes \cite{daza}. The computational power of the fixed infrastructure devices are greater than the nodes i.e, the nodes are less computational because of their restrictions on the amount of energy utilization. Denial of service, eavesdropping, interception and routing 
attacks are major threats to security in MANETS.\cite{perrig} \cite{perrig2}.
The advent of IOT devices and the wide usage of the electronic devices raised concerns over these security issues. There are 
several factors like power dissipation, area and cost that affect embedded applications which implement a full-fledged 
cryptographic environment. Because of these factors the concept of lightweight cryptography emerged. Less amount of memory space is utilized by lightweight cryptography. As a result of this, the amount of gate equivalent count is less and eventually leads to an efficient hardware implementation. Nowadays Public key cryptosystem(PKC) is used widely in many applications. 
The main concept of PKC is to provide key exchange services and digital signatures for authentication purpose. In PKC, the 
concept of public key management is to be handled carefully to avoid the wide range of attacks. In PKC, an authority looks 
after all the public parameters and generates them and will provide the details when necessary to a particular user. The 
authority might also generate the private keys too. But as it is a third party, there are cases where the authority might 
misuse its power, therefore the trusted authority will generate only partial private key \cite{pkc}. In our proposal, the 
nodes of the MANET hold a partial share of the MANET secret and they use that share to generate a secret key using Diffie - 
Hellman key exchange to communicate among them

\section{Preliminaries}

\subsection{Shamir Secret Sharing}
Blakley \cite{blakley} and Shamir \cite{shamir} are the first to introduce secret sharing techniques. In general a secret 
sharing scheme contains a dealer $D$ and a set \ $U= \{ {u_1,u_2,\dots,u_n} \} $ of $n$ users. The dealer has a secret 
$s$ and a share $s_i$ of which is privately distributed to user $u_i, 1\le i \le n$. A valid subset  $u$ ( for : $u \subset 
U $) of $t$ number of users holding valid shares can reconstruct the secret $s$, where $t$ is minimum number of users or threshold value and $(t,n)$ scheme is called as threshold  access structure \cite{shamir}.
In threshold cryptography, the private key $s$ is shared among $n$ participants using a (t,n) threshold access structure 
with the help of a secret sharing scheme, with each participant $u_i$ having a partial share $s_i$ \cite{zhou}. For example, 
in public key cryptography(PKC), let the public key be $pk$ and the corresponding private key be $s$. If a user has 
encrypted a message using the public key $pk$ then to decrypt the message, at-least $t$ out of $n$ nodes are needed to 
decrypt the message.

Shamir's secret sharing scheme realizes a (t, n) threshold access structures by means of polynomial interpolation. Let $Z_q$ 
be a finite field with $q > n$ and let $s \in Z_q$ be the secret. A polynomial P(x) is taken by the dealer having degree $t-1$, 
here the actual secret is $s$ which is a constant term of $P(x)$ and all other coefficients $a_i$ for $i\in {1,2,3,\dots,t-1}$ are selected from 
$Z_q$  uniformly and independently at random. That is,
$P(x)=s+ \sum_{\substack{i=1}}^{t-1}a_i*x^i$.
Every party or user $u_i$ is associated to a field element $b_i$ publicly for $i\in {1,2,3,\dots,n}$. The different field elements are associated with different parties or users. The dealer privately sends to party $u_i$ the value $[s]_i = P(b_i), for\ i = 1,2,\dots, n$.This 
scheme realizes $(t, n)$ threshold access structure. The parties agreeing to recover secret $s$ is ${u_1, \dots, u_t }$. The secret $s$ can be obtained as $\sum\limits_{i=1}^{t} [s]_i 
\prod\limits_{m=0,m\neq j}^t \frac{x_m}{x_m-x_j}$. The number of parties less than t cannot interpret any information regarding the secret $s$. We used basic Shamir's secret sharing to share the 
key among the MANET nodes, however depending on the application different variations can also be used like the one proposed 
by Abdul et. al. \cite{BASIT2017} and \cite{singh2016sequetial}.

\subsection{Elliptic Curve Cryptography \cite{neal} and Diffie Hellman Key Exchange}
Elliptic curve cryptosystems were first proposed independently by Neil Koblitz and Victor Miller in 1985. Elliptic curve 
cryptography (ECC) is a public key encryption technique that is  based on the algebraic structure of elliptic curves over 
finite fields. In comparison to non-Elliptic curve cryptography, Elliptic curve cryptography makes use of relatively smaller keys to provide the same level of security. The discrete logarithmic defined using a group of points on an elliptic curve over a finite field is the hardest among the remaining groups. The security of a cryptosystem relies on the hardness of this discrete logarithmic problem.
 \cite{victor}.

\subsubsection{Elliptic Curve Discrete Logarithm Problem (ECDLP) }
ECDLP is the problem of finding the integer s, given a rational point Q on the elliptic curve E and the value of s*Q. Elliptic curve cryptosystems are dependent on the level of difficulty of ECDLP. If an attacker can solve the ECDLP then he will be able to break the system, but ECDLP is much harder than Discrete Logarithm Problem(DLP) in finite fields. The strongest techniques generally used to solve DLP in finite fields, like the index calculus method, Shank's baby step giant step algorithm, the Pohlig - Hellmann method and Pollard's $\rho$ method don't work in solving ECDLP problem. Most of them work only if the group order is divisible by some large prime. In 1993, Menezes, Okamoto and Vanstone reduced the ECDLP to DLP on $F^*_(q^k)$. This method works only for supersingular curves i.e curves of the form $ y^2 = x^3 + a x$ when the characteristic $p$ of $F_q \equiv -1(mod  \; 4)$ and also curves of the form $y^2=x^3+b$ when $p \equiv -1(mod \; 3)$ i.e curves for which k is small. This method known as MOV method is only useful for a small class of elliptic curves but most elliptic curves are non supersingular.

\subsubsection{The Diffie Hellman Key Exchange using ECC}
Symmetric Encryption of data requires the transfer of secret key from the sender to receiver without anyone intercepting the key,  which is the most challenging task \cite{dh}. Diffie-Hellman algorithm made the transfer and the generation of similar keys at two ends secretly successful.   The idea of the Diffie-Hellman algorithm was coined by Diffie and Martin Hellman in the year 1976. The key goal behind the development of this algorithm was not mere encryption and decryption of data, but it completely revolved around the generation of a similar private cryptographic key at two ends so that the transferring of a key.\cite{dhk}.  The algorithm may lag behind due to its speed, but the power of the algorithm is the ultimate reason behind its popularity among key generation techniques.  The following procedure generates a common secret key between two parties A and B using ECC.
\begin{itemize}
\item A and B agree on a Finite field $F_q$ and an elliptic curve $E$ defined over $F_q$.
\item They also choose a public random base point $P$ that belongs to $E$.
\item A chooses a random secret integer $a$, computes $a \cdot P \in E$ and sends it to B.
\item B chooses a secret random integer $b$, computes $b \cdot B \in E$ and sends it to A.
\item A computes the secret key $s = a(b \cdot P)$ and B computes the secret key $s = b(a \cdot P)$. so they both have the common secret point s.
\item If the attacker gets hold of $P, a \cdot P, b \cdot P$, then there is no easy way to calculate $abP$ because of ECDLP.
\end{itemize}

\subsection{Lightweight Cryptography}
The type of cryptography which deals with the integration of cryptographic primitives, ciphers and various techniques utilizing a small computing power and deliver high standards of security is known as lightweight cryptography. To design lightweight algorithms, the computational complexity and the demands on hardware and other limitations of the device are to be analyzed. In this paper we considered Tiny Encryption Algorithm (TEA), which is one of the fastest and most efficient lightweight algorithm.

\subsubsection{Tiny Encryption Algorithm(TEA) \cite{pretea}}
TEA is a symmetric Feistel type cipher created by David Wheeler and Roger Needham of Cambridge University, which come under a specialized class possessing iterated block ciphers. The ciphertext is generated from the plain text by repeatedly applying the same transformation or round function. In this form of cipher, there are two stages of encryption process. The first stage is the application of a round function F on the first part with the use of a subkey  and the output of this stage is XOred with the remaining part. The above mentioned pattern is followed for each round except the final round where the swapping is not applied. A double shift operation in TEA results in the mixing of key and all bits containing the data repeatedly. The 128-bit key K is split into four 32-bit blocks (K = (K[0], K[1], K[2], K[3]) by using the key shift algorithm.
 
\paragraph{ENCRYPTION OF TEA} \cite{encdec}
A plain text P of 64 bit length will be splitting into 2 equal halves i.e 32 bit each. The encryption of one half is done with the other half by processing it in 64 rounds. Later this is joined and eventually leads to the formation of cipher text block( C = Left[64], Right[64]). A 128 bit key K is divided in to four parts ( K =K[0], K[1], K[2], K[3]).
\begin{itemize}
\item A sub key k[i], Left[i-1] and Right[i-1] act as inputs to each round i which are the output from the previous round.
\item The sub keys K[i] are different from K and from each other.
\item The constant delta = $(\sqrt{5} - 1) * 2 ^{31} = 0x9E3779B9$.
\item The round function slightly differs from a classical Fiestel cipher structure in that integer addition modulo $ 2^{32}$ is used instead of exclusive-or as the combining operator.
\end{itemize}

\begin{figure}
\begin{center}
\includegraphics[scale=0.5]{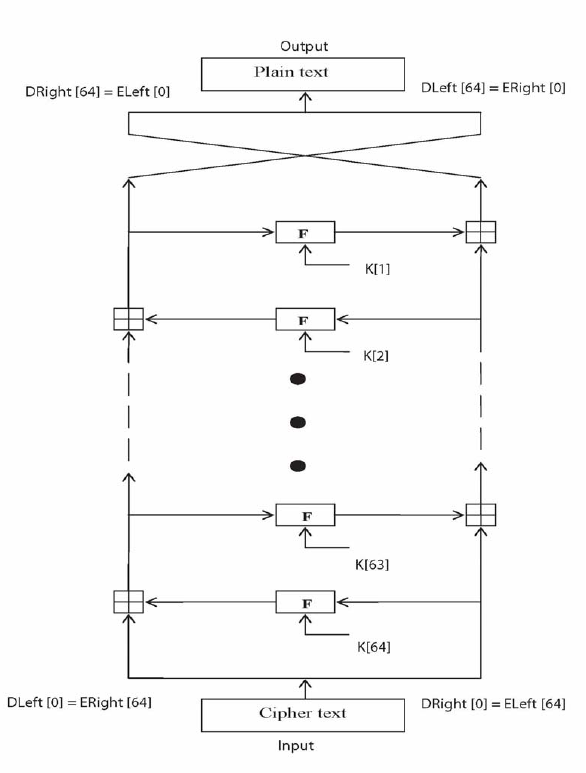}
\caption{TEA Encryption Structure}
\end{center}
\end{figure}

The structure of TEA encryption algorithm is depicted in figure 1 and the symbol $\boxplus$ denotes integer addition modulo $2^{32}$. The round function F, consists of the key addition, bit wise XOR and left and right shift operations. The output (Left[i+1],Right[i+1]) of the $i^{th}$ cycle of TEA with respect to the input (Left[i],Right[i]) is described as follows :
\begin{itemize}
\item[•] $Left[i+1] = Left[i] \; \boxplus \; F(Right[i], \; K[0,1], \; delta[i])$ 
\item[•] $Right[i+1] = Right[i] \; \boxplus \; F(Right[i+1], \; k[2,3], \; delta[i])$
\item[•] $delta[i] = (i+1)/2 * delta $
\end{itemize}
The round function F, is denoted by \\ F(M, K[j,k], delta[i]) = ((M $\ll$ 4) $ \boxplus$ K[j]) $\oplus$ (M $\boxplus$ delta[i])$\oplus$ ((M $\gg$ 5)$\boxplus$ K[k]), where j,k ranges from 1 to 64.
The round function structure is same for each round but takes parameters from the round sub key K[i]. It uses simple key schedule algorithm, the 128-bit key K is divided into four 32-bit blocks K = {K[0],K[1],K[2],K[3]). The odd rounds used the keys K[0] and K[1]  and the even rounds use the keys K[2] and K[3]. Figure 2 shows the internal details of the $i^{th}$ cycle of TEA.
\begin{figure}
\begin{center}
\includegraphics[scale=0.5]{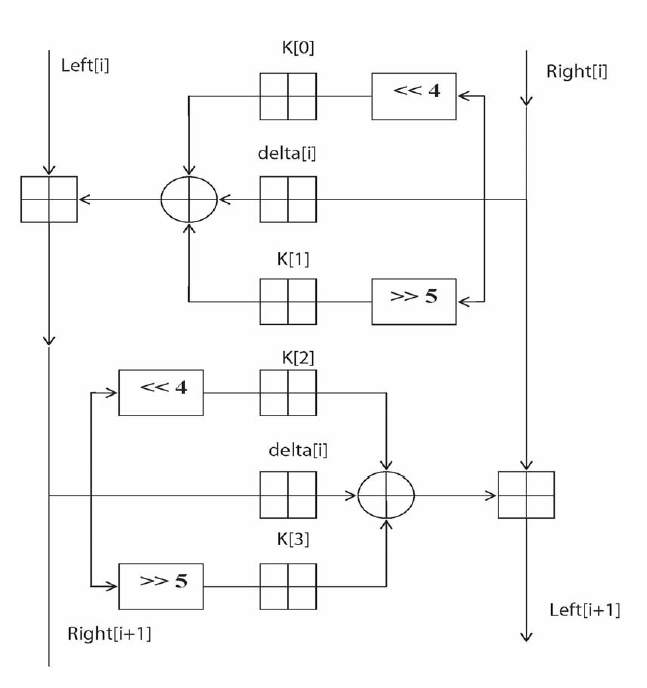}
\caption{General View of TEA in its i-th Cycle}
\end{center}
\end{figure}

\paragraph{DECRYPTION OF TEA} 
The decryption process initiates with accepting the cipher text as input followed by the reverse application of the K[i] subkeys. This process is similar to the encryption process.The Decryption is similar to encryption which takes cipher text as input and the sub keys K[i] are iapplied in the reverse order. The intermediate value of the decryption process is equal to the corresponding value of the encryption process with the two halves of the values swapped. For example, if the output of the $i^{th}$ encryption round is \\
$ELeft[i] || ERight[i]$  (ELeft[i] concatenated with ERight[i]).\\
Then the corresponding input to the $(64-i)^{th}$ decryption round is \\
$DRight[i] || DLeft[i]$   (DRight[i] concatenated with DLeft[i]).\\
Once the last iteration of the encryption process is completed then the two halves of the output are exchanged, so that the cipher text is $ERight[64] || ELeft[64]$ and the final cipher text C is the output of that round. Figure 3 represents the structure of the TEA decryption process.

\begin{figure}
\begin{center}
\includegraphics[scale=0.5]{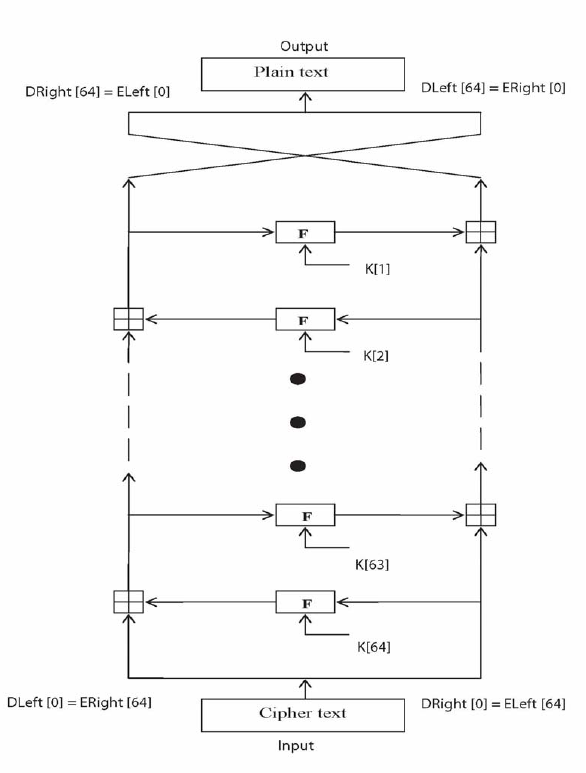}
\caption{The structure of TEA decryption process}
\end{center}
\end{figure}

\section{Proposed System}
In our Proposed system, we show how the nodes of MANET can communicate securely using lightweight cryptography.
An additive group $G$ being generated by some element $Q$ having prime order $q$, and is hard under the assumption of discrete logarithmic problem are made public. The public key is given as $PK=s*Q$. 
We use a collision-resistant hash function $h:\{0,1\}^*\rightarrow Z_q$. All the values $G,q,Q,PK$ are made public. The nodes exchange the private keys by using the Diffie hellman key exchange. Our proposed scheme consists of the phases $Initial\ Setup$, $Key\ Distribution$, $Computing \ a \ secret \ Key$ and  $Secure\ communication$.

\subsection{Initial Setup}
The very first step is to make the MANET as de-centralized as possible, i.e., the role of the dealer has to be played by nodes themselves.

Let $N$ be the initial set of k founding nodes in the MANET, $t$ be the threshold for $t\leq k$. Every founding node $n_i \in N$  chooses a bi-variate  polynomial, symmetric in $x,z$ with the degree of $x\ and\ z$ being $t-1$, where $t$ is the threshold number. Every node $n_i \in N$ obtains its partial secret $s_i$ by implementing Shamir secret sharing scheme. The protocol also supports new nodes joining the network, by obtaining their respective partial share. The complete step by step procedure is described in the next section $Key\ Distribution$

\subsection{Key Distribution}

Every node $n_i \in N$ which is part of the MANET receives partial share $s_i$ of the actual  secret $s$. This is achieved using the following protocol.
\begin{enumerate}
\item Every founding node $n_i \in N$ chooses a bi-variate polynomial $F_i(x,z) \in Z_q[x,z]$, symmetric in $x,z$ and the degree of $x\ and\ z$ is at most $t-1$, where $t$ is the threshold value. These polynomials implicitly define a polynomial $ F(x,z) = \Sigma_{n_i\in N}F_i(x,z)$. Let us denote $f_{i,0} = F_i(0,0)$ the constant term of polynomial $F_i(x,z)$ and $s=\sum\limits_{i=1}^{k}f_{i,0}=F(0,0)$ the constant term of $F(x,z)$.
\item Every node $n_i \in N$ computes and secretly sends to all other  nodes $n_j \in N$ the univariate polynomial $F_{ij}(x) = F_i(x,h(n_j)$, along with the value $Y_i = F_{i}(0,0)*Q$ (where Q is the generator of Group G). 
\item Finally every node $n_j$ has values received from other k-1 nodes and also has it's own value $F_{jj}(x,h(n_j))$ with it. Then every node $n_j \in N$ computes  $S_j(x)=\sum\limits_{j=1}^{k} F_{ij}(x) =\sum\limits_{j=1}^{k} F_i(x,h(n_j) = F(x,h(n_j))$ 
\item Now every node $n_jn N$ has a share  $s_j=S_j(0) = F(0,h(n_j)$ of implicit MANET secret $s = F(0,0)$ and a secret univariate polynomial of its own  $S_i(x)$.
\item The public key $PK$ of the MANET is made public by each node in MANET. This public key is computed by each node in MANET with the help of the information acquired from the rest of the nodes in the initial setup. The public key is given as  $PK=\Sigma_{i\in N}F_i(0)*Q=s*Q$

\end{enumerate}

The MANET secret function $F(x,z)=\sum\limits_{i=1}^{k} F_i(x,z)$ and MANET secret key  $s=F(0,0)$ are safe and hidden as 
they are implicit and not known to any particular node. The reconstruction of the secret takes place if any only if minimum of $t$ nodes possessing the partial secret of the actual MANET secret come together. Key distribution process is depicted in Figure 4. The shares of the nodes can be frequently updated by using the concept of 
proactive secret sharing as discussed in \cite{NCK2}.
\begin{figure}
\begin{center}
\includegraphics[height=16cm,width=16cm]{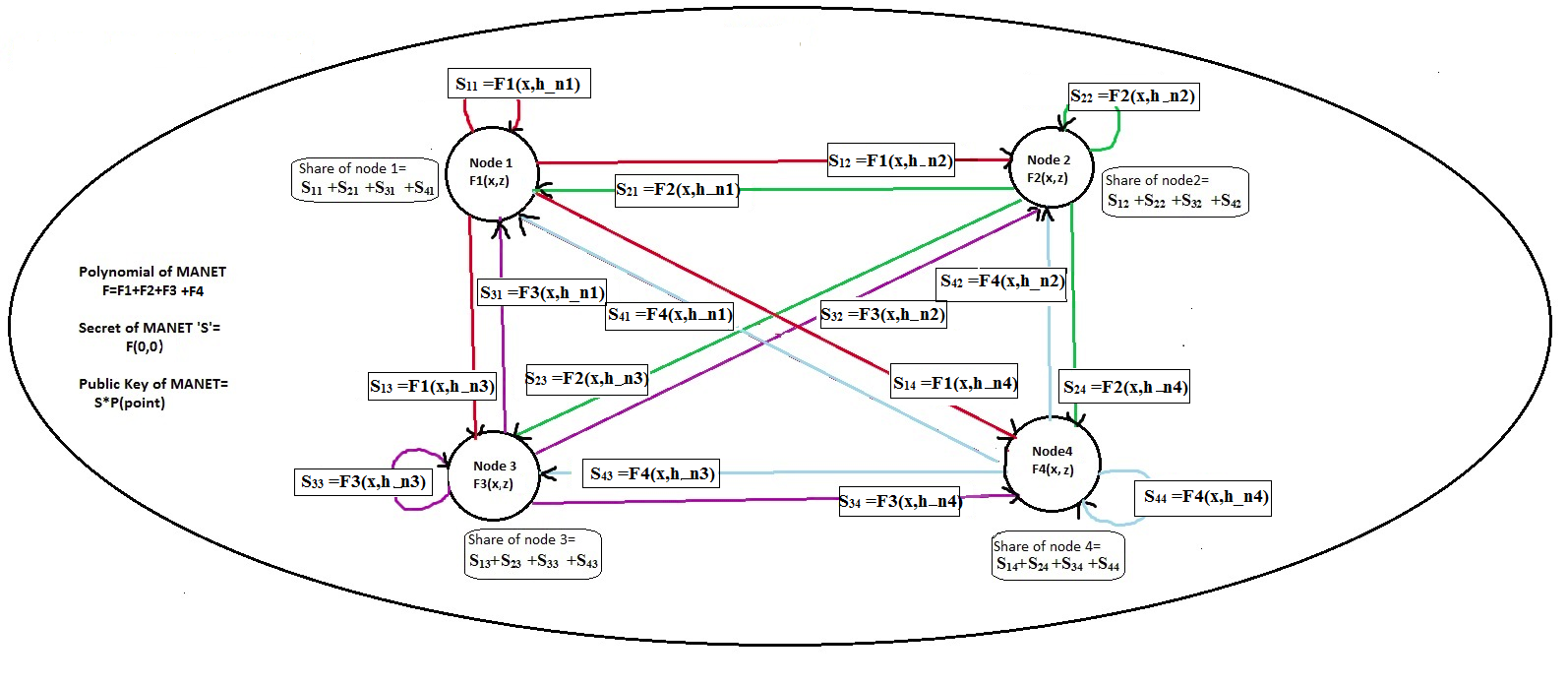}
\caption{Key Distribution process}
\end{center}
\end{figure}

\subsection{Computing a Secret Key}
If the nodes $ n_i, n_j \in N$ want to communicate securely then they need a common secret key, which is used for both 
encryption and decryption. The nodes follow the Diffie Hellman Key exchange algorithm discussed in section 2.2.2 to compute 
the secret key. They use the following protocol:
The General Algorithm for Diffie Hellman is as follows:
\begin{enumerate} 
\item $n_i$ and $n_j$ choose random numbers $a,b \in Z_q$.
\item $n_i$ sends a point $ A= a*Q$ (where Q is the generator of Group G) and $n_j$ sends $B=b*Q$ to $n_i$
\item $n_i$ computes a point $R=a*B = a*b*Q$ and $n_j$ computes $R = b*A = a*b*Q$. 
\item Both $n_i$ and $n_j$ have same secret point $R = E(R_x,R_y)$ where $R_x and R_y$ represent x and y coordinate of the 
point R. The secret key $sk$ is computed as $ sk = R_x + R_y$.

\end{enumerate}

\subsection{Secure Communication }
Once the nodes $n_i,n_j \in N$ have a common secret key $sk$ then they communicate securely using TEA algorithm discussed in 
section 2.3.1.

\section{Example}
The MANET setup is similar to Chaitanya et al. discussed in \cite{NCK}.

\begin{itemize}
\item \textbf{Setup}
		\item Let the intial set of nodes $N = \{N_1,N_2,N_3,N_4\}$    \\ No. of Nodes = l = 4
		\item Public Parmeters : \\
		An additive group G is selected with the prime order q = 83.\\
		- The curve used is $E(F_{83}) : y^2 = x^{3} +  1 $ \\
		- The Generator is Q = E(38,50)\\
 		- Let t = 3.
         
         \item A collision resistant hash function - HTR
		
	\end{itemize}

	\begin{itemize}
		\item Each node selects a random bivariate polynomial in GF(83)\\
		$  N1 = 3x^2z+3z^2x+8xz+5z+5x+5$\\
		$ N2 = 5x^2z+5z^2x+3xz+8z+8x+9$\\
		$N3 = 8x^2z+8z^2x+5xz+3z+3x+6$\\
		$N4 = 2x^2z+2z^2x+4xz+8z+8x+4$
		\item The polynomial implicitly defined by all the nodes is \\
		F(x,z) = $ N_1 + N_2 + N_3 + N_4 $\\
		       = $18x^2z + 18xz^2 + 20xz + 24x + 24z + 24$ 
		\item The MANET secret $s$  F(0,0) = 24.
	\end{itemize}

\begin{itemize}
		\item Each node secretly sends to each of other founding nodes the univariate polynomial $F_{ij} = F_i(x,h(N_j)) $.
		\item The hash function used is given below:\\
			  def HTR(id,p): \\
    		  h = int(hashlib.sha224(str(id)).hexdigest(),16)\\
    		  val = mod(h,p)\\
    		  return val
		\item The hash values of the nodes are \\
		$ h_{n1} = HTR('Node1',k) = 21 $ \\
		$ h_{n2} = HTR('Node2',k) = 57 $ \\
		$ h_{n3} = HTR('Node3',k) = 63 $ \\
		$ h_{n4} = HTR('Node4',k) = 31 $ 
		\end{itemize}
		
		\begin{itemize}
		\item \textbf{Share Distribution}
		\item Each node sends the following values to other Nodes :
		\item  N1 also includes $Y_1$ = 5 * Q = (18,43) \\
		$N_{11} = 63 x^{2} + 2 x + 27 $\\
		$N_{12} = 5 x^{2} -  x + 41 $\\
		$N_{13} = 23 x^{2} + 49 x + 71 $\\
		$N_{14} = 10 x^{2} + 65 x + 77 $
		\item N2 also includes $Y_2$ = 9 * Q = (57,41) \\
		$N_{21} = 22 x^{2} + 35 x + 11 $\\
		$N_{22} = 36 x^{2} + 73 x + 50 $\\
		$N_{23} = 66 x^{2} + 39 x + 15 $\\
		$N_{24} = 72 x^{2} + 9 x + 8 $	
		\item N3 also includes $Y_3$ = 6 * Q = (68,64) \\
		$N_{31} = 2 x^{2} + 67 x + 69 $\\
		$N_{32} = 41 x^{2} + 52 x + 11 $\\
		$N_{33} = 6 x^{2} + 32 x + 29 $\\
		$N_{34} =  - x^{2} + 44 x + 16 $
		\item N4 also includes $Y_4$ = 4 * Q = (48,55) \\
		$N_{41} = 42 x^{2} + 61 x + 6 $\\
		$N_{42} = 31 x^{2} + 11 x + 45 $\\
		$N_{43} = 43 x^{2} + 64 x + 10 $\\
		$N_{44} = 62 x^{2} + 62 x + 3$			
		\item Then from the received values all the nodes calculate their secret univariate polynomial.
		\item $S_1(x) =46 x^{2} -  x + 30 $ 
		\item $S_2(x) = 30 x^{2} + 52 x + 64 $
		\item $S_3(x) = -28*x^2 + 18*x - 41 $
		\item $S_4(x) = -23*x^2 + 14*x + 21 $ 
		
		\end {itemize}
		
		\begin{itemize}
		\item The public key, PK = s * Q \\  $  \hspace{3cm} $   = 24 * E(38,50) = E(11,81)
		\item PK should also be equal to $Y_1 + Y_2 + Y_3 + Y_4$\\
		=E(18,43)+E(57,41)+E(68,64)+E(48,55)\\ = E(11,81)
		\item Each node computes its share from $S_i(0)$.\\
		The shares of the nodes are \\ $S_1 = 30, S_2 = 64 ,S_3 = 42, S_4 = 21$
\item If $n_1$ and $n_2$  wants to communicate securely then they compute a common secret key using their shares.
\item $n_1$ sends $S_1 * Q = 30 * E(38,50) = E(35,31)$ to $n_2$.
\item $n_2$ sends $S_2 * Q = 64 * E(38,50) = E(50,70)$ to $n_1$.
\item Both $n_1$ and $n_2$ computes the same secret point. At $n1$,
$30 * E(50,70) = E(6,47)$ at $n_2$,$ 64 * E(35,31) = E(6,47)$.
\item Now both $n_1$ and $n_2$ have a secret key 53, which they use for encryption and decryption.
\item If $n_1$ wants to send a message "hello" to $n_2$ using TEA,
then it sends the cipher text ['0x95c88604', '0x1745f2d7'].
\item Then $n_2$ decrypts the cipher text to message "hello" using the key 53.

\end{itemize}

\section{Security Analysis}
The TEA algorithm is prone to key equivalence attack as discussed below:{Key equivalence Attack}:This attack on TEA is done 
by using Known plain/cipher text pairs of any unknown key (K). The general idea 
of the attack is that the key in TEA is divided into four 32 bits (key=K[0], K[1],K[2],K[3]) and for one half of the plain 
text K[0],K[1] are used and K[2],K[3] for another half. So, if we find the value of K[0] then we can find K[1] (same with 
other half) because$ R[i+1] =L[i]+(((R[i]<<4)+K[0])  \oplus((R[i]>>5)+K[1])\oplus(R[i]+Delta)))$. As all values are known  
we need to guess the value of K[0] starting from zero.and calculate k[1] for the first and second plain text/cipher text 
pairs. If the these two values do not match then increment the value of K[0] and proceed form the beginning. If they match 
then verify whether this guess of K[0] is correct by checking K[1] for other plain text/cipher text pairs. If the value of 
K[1] matched every time then this is the correct guess for K[0] and the key is found. But we can avoid this attack in MANETs 
by adopting proactive secret sharing techniques, where the nodes periodically update their shares thus not giving enough 
time to the attacker to find the equivalent keys.

\section{Conclusion}
In this paper, we proposed the use of lightweight cryptography algorithm in MANETs. As MANETs have less computation power, 
they can not use traditional cryptography algorithms because of heavy computations. In our MANET setup, first all the nodes 
have a share of MANET private key  and they use this share to generate a secret key using Diffie Hellman key exchange. Then 
they use the  secret key with the TEA algorithm to encrypt/decrypt messages. We also discussed that the general key 
equivalence attack of TEA can be avoided by adopting to proactive secret sharing.

\bibliographystyle{elsarticle-num}

\vspace{2cm}

\section*{Authors}
\noindent {\bf N Chaitanya Kumar} received M.Tech from JNTU Hyderabad, and he did Bachelor degree in computer science. 
Currently, he is pursuing his PhD in Computer Science from the University of Hyderabad. His research interests include Information security, Cryptography in MANET.\\

\noindent {\bf Abdul Basit} received Master of computer application from Jamia Hamdard University New Delhi. He did Bachelor of Science in Information technology from SMU Gangtok. Currently, he is pursuing his PhD in Computer Science from the University of Hyderabad. His research interests include Information security, Cryptography, and Cyber security.\\

\noindent {\bf Priyadarshi Singh} received M.Tech from IIT(ISM) Dhanbad. He did Bachelor degree in Information Technology. Currently, he is pursuing his PhD in Computer Science from the University of Hyderabad. His research interests include Cryptography, Public key infrastructure.\\

\noindent {\bf V. Ch. Venkaiah} obtained his PhD in 1988 from the Indian Institute of Science (IISc), Bangalore in the area 
of scientific computing. He worked for several organisations including the Central Research Laboratory of Bharat Electronics, Tata Elxsi India Pvt. Ltd., Motorola India Electronics Limited, all in Bangalore.
He then moved onto academics and served IIT, Delhi, IIIT, Hyderabad, and C R Rao Advanced Institute of Mathematics, 
Statistics, and Computer Science. He is currently serving the Hyderabad Central University. He is a vivid researcher. He 
designed algorithms for linear programming, subspace rotation and direction of arrival estimation, graph coloring, matrix symmetriser, integer factorisation, cryptography, knapsack problem, etc.\\

\end{document}